\documentclass[aps,prl,twocolumn,superscriptaddress]{revtex4-2}

\usepackage{graphicx}
\usepackage{xcolor}

\begin{document}

\title{Self-Similar Liquid Lens Coalescence}

\author{Michiel A. Hack}
\email{m.a.hack@utwente.nl}
\affiliation{Physics of Fluids Group, Faculty of Science and Technology, University of Twente, P.O. Box 217, 7500 AE Enschede, The Netherlands}

\author{Walter Tewes}
\affiliation{Physics of Fluids Group, Faculty of Science and Technology, University of Twente, P.O. Box 217, 7500 AE Enschede, The Netherlands}

\author{Qingguang Xie}
\affiliation{Department of Applied Physics, Eindhoven University of Technology, P.O. Box 513, 5600 MB Eindhoven, The Netherlands}

\author{Charu Datt}
\affiliation{Physics of Fluids Group, Faculty of Science and Technology, University of Twente, P.O. Box 217, 7500 AE Enschede, The Netherlands}

\author{Kirsten Harth}
\affiliation{Physics of Fluids Group, Faculty of Science and Technology, University of Twente, P.O. Box 217, 7500 AE Enschede, The Netherlands}
\affiliation{Institute of Physics, Otto von Guericke University, 39106 Magdeburg, Germany}

\author{Jens Harting}
\affiliation{Helmholtz Institute Erlangen-N\"urnberg for Renewable Energy (IEK-11), Forschungszentrum J\"ulich, F\"urther Str. 248, 90429 Nuremberg, Germany}
\affiliation{Department of Applied Physics, Eindhoven University of Technology, P.O. Box 513, 5600 MB Eindhoven, The Netherlands}

\author{Jacco H. Snoeijer}
\affiliation{Physics of Fluids Group, Faculty of Science and Technology, University of Twente, P.O. Box 217, 7500 AE Enschede, The Netherlands}

\date{\today}

\begin{abstract}
A basic feature of liquid drops is that they can merge upon contact to form a larger drop. 
In spite of its importance to various  applications, drop coalescence on pre-wetted substrates has received little attention. 
Here, we experimentally and theoretically reveal the dynamics of drop coalescence on a thick layer of a low-viscosity liquid. 
It is shown that these so-called ``liquid lenses'' merge by the self-similar vertical growth of a bridge connecting the two lenses. 
Using a slender analysis, we derive similarity solutions corresponding to the viscous and inertial limits. 
Excellent agreement is found with the experiments without any adjustable parameters, capturing both the spatial and temporal structure of the flow during coalescence. 
Finally, we consider the crossover between the two regimes and show that all data of different lens viscosities collapse on a single curve capturing the full range of the coalescence dynamics. 
\end{abstract}


\maketitle

The coalescence of liquid drops is an important part of many industrial processes, such as inkjet printing and lithography \cite{Wijshoff2018, Winkels2011}. 
It is also ubiquitously observed in nature, for example in the formation of rain drops and the self-cleaning of plant leaves \cite{Grabowski2013, Wisdom2013, Watson2014}. 
Coalescence, therefore, has been the focus of many studies, primarily for spherical drops \cite{Eggers1999, Duchemin2003, Aarts2005, Thoroddsen2007, Paulsen2011}, but also for drops on a solid substrate \cite{Ristenpart2006, Narhe2008, Lee2012, HernandezSanchez2012, Eddi2013}. 
In contrast, little work exists on the coalescence of drops on liquid substrates \cite{Burton2007,Delabre2010}, despite its importance for emerging applications such as fog harvesting \cite{Sokuler2009, Anand2012}, anti-icing \cite{Kim2012}, wet-on-wet printing \cite{Hack2018}, enhanced oil recovery \cite{Wasan1978, Wasan1979}, emulsions \cite{Shaw2003, Kamp2016, Keiser2017}, and wetting of lubricant-impregnated surfaces \cite{Smith2013}.

\begin{figure}
\includegraphics{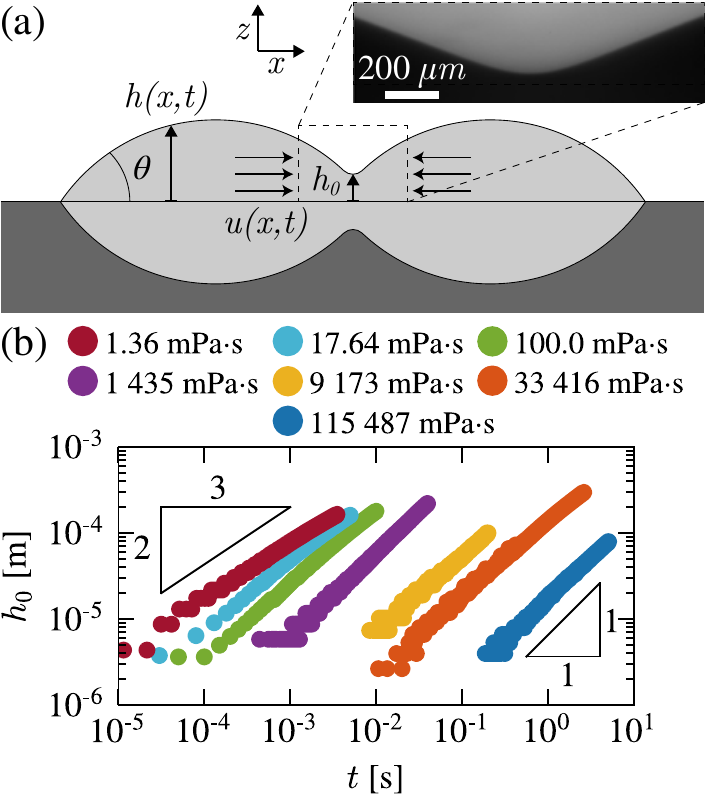}%
\caption{\label{fig1} (a) Schematic view of two coalescing liquid lenses connected by a bridge of height $h_0(t)$. The lenses float on a pool with a  depth that is much larger than the size of the lenses. The zoomed region shows a typical snapshot of the bridge region. (b)  Measurements of the bridge height $h_0$ as a function of time $t$ for several viscosities. Two distinct power laws are identified.}
\end{figure}

The dynamics of coalescence are strongly affected by the geometry of the drops. 
Drops on a solid substrate (spherical caps, \cite{Ristenpart2006, Narhe2008, Lee2012, HernandezSanchez2012, Eddi2013}) merge differently than freely suspended drops (axisymmetric spheres, \cite{Eggers1999, Duchemin2003, Aarts2005, Thoroddsen2007, Paulsen2011}), with different scaling exponents for the growth of the bridge between the drops. 
This is in contrast to the coalescence of drops floating on a liquid substrate (Fig.~\ref{fig1}a); such drops are referred to as ``liquid lenses"  \cite{Langmuir1933, deGennes2004}. 
For coalescing lenses, the growth of the bridge width based on a top-view experiment was found similar to that of axisymmetric drops \cite{Burton2007}, which is surprising since, geometrically, liquid lenses are spherical caps. 

\begin{figure*}
\includegraphics{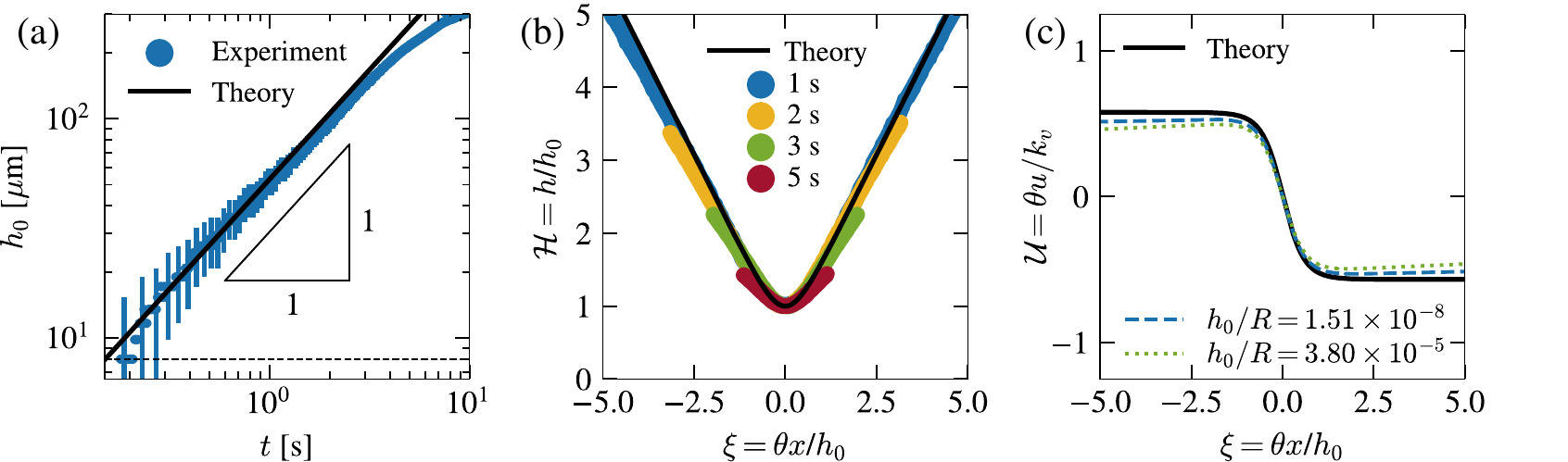}%
\caption{\label{fig2} Coalescence in the viscous regime. (a) Height of the bridge $h_0$ as a function of time after contact $t$ (mineral oil lenses, $\theta = 33^\circ$, $\eta = 115~487$ mPa$\cdot$s, initial height $\approx~0.5$ mm). The solid line is the prediction from (\ref{eqV}). The error bars are only shown for one in every ten datapoints for clarity. The horizontal dashed line indicates the resolution limit. (b) Rescaled experimental profiles at different times, $\mathcal{H} = h(x,t)/h_0(t)$ versus $\xi = x\theta/h_0(t)$. The collapse of the profiles indicates self-similar dynamics. The solid line is the similarity solution obtained from (\ref{eqMassDimensionless}, \ref{eqMomentumDimensionlessViscous}). (c) Rescaled velocity profile. The solid line is the similarity solution. The colored lines are numerical simulations for different values of $h_0/R$.}
\end{figure*}

In this Letter, we study the coalescence dynamics of liquid lenses in terms of the vertical bridge growth $h_0(t)$ (defined in Fig.~\ref{fig1}a), and reveal a strong departure from the coalescence of axisymmetric drops. We first experimentally establish the initial dynamics of coalescence of drops of varying viscosity from the side-view perspective, identifying two distinct regimes -- one dominated by viscosity and the other by inertia. 
Subsequently, we develop a fully quantitative slender description for each of these regimes based on the self-similar nature of coalescence. 
In the spirit of recent work on spherical drops \cite{Paulsen2011, Xia2019}, we identify the master curve for all data, including the crossover between the two regimes. 
Unlike for any other coalescence problem, however, the master curve here is obtained without any adjustable parameter. 

\emph{Coalescence dynamics.}---Two small drops are placed on a deionized water surface (MilliQ, Millipore Corporation) kept in a large container.
The lenses consist of mineral oils (RTM series, Paragon Scientific Ltd.), with viscosities between $\eta =$~18~mPa$\cdot$s and 115~Pa$\cdot$s and surface tension $\gamma =$~34~mN$\cdot$m$^{-1}$ (measured by the pendant drop method \cite{Hansen1991}). 
Additionally, we use dodecane lenses (Sigma-Aldrich, $\eta =$~1.36~mPa$\cdot$s, $\gamma =$~25~mN$\cdot$m$^{-1}$). 
These liquids float on the water surface since their densities ($\rho =$~850~kg$\cdot$m$^{-3}$ for mineral oil and $\rho =$ 750~kg$\cdot$m$^{-3}$ for dodecane) are lower than the density of water ($\rho =$~997~kg$\cdot$m$^{-3}$). 
Both liquids have a negative spreading parameter, and thus form lenses with small but finite contact angles $\theta = 26^\circ$ to $37^\circ$ \cite{deGennes2004}. 
Since the contact angle of the oil-water interface is within $5^\circ$ of the aforementioned values, we regard the lenses as being top-down symmetric. 

We image the coalescing lenses from the side using a high-speed camera (Photron Nova S12) equipped with a microscopic lens (Navitar 12X zoom lens).
In order to obtain a sharp image of the oil-air interface of the liquid lens, the container of the pool is filled such that a convex meniscus forms at the edges of the container. 
Frame rates between 250 frames/s and 100~000~frames/s are used depending on the timescale of coalescence, with resolutions in the range of  1.3--5.3~$\mu$m/pixel. 
A typical snapshot of the bridge region is shown in the zoomed region in Fig.~\ref{fig1}a.

The experiment is performed as follows; two pendant drops with volume $V =2.5$~$\mu$L are formed on two identical blunt-ended metal needles using a syringe pump (we have verified that drop size does not affect the initial coalescence dynamics).
Using a linear translation stage, the drops are gently brought into contact with the water pool and subsequently form lenses of radius $R\approx2.5$~mm. 
The lenses are left to equilibrate for a moment before the syringes are gently removed. 
Capillary interactions drive the lenses toward each other and they coalesce upon first contact. 
We define $t=0$ as the first frame where the bridge connecting the two lenses is visible, and $h = 0$ at the surface of the pool.
The velocity of the approaching lenses is orders of magnitude smaller than the velocity of the bridge growth. 

The experiments reveal that the coalescence of liquid lenses is governed by a self-similar power-law growth of the bridge that connects the two drops. 
Figure~\ref{fig1}b shows the minimum bridge height $h_0$ as a function of time after contact $t$ for coalescing lenses of different viscosities. 
We clearly distinguish two regimes: a nonlinear regime for small viscosities where $h_0 \propto t^{2/3}$, and a linear regime where $h_0 \propto t$ for high viscosities. 
These exponents are typical for pinch-off and coalescence of spherical caps on a solid substrate \cite{HernandezSanchez2012, Eddi2013, Eggers2008, Day1998}. 
These growth dynamics, however, are different from those of spherical drops and of those observed for lenses in top-view \cite{Burton2007}. To further investigate this, we now first focus on the case of viscous coalescence. 
Figure~\ref{fig2}a shows the temporal evolution of the bridge, which grows at constant velocity. 
The bridge velocity decreases when $h_0$ becomes of the order of the lens size, due to the finite height ($\approx$~0.5~mm) of the lens.
The spatial structure of coalescence is revealed in Fig.~\ref{fig2}b, where we compare the shape of the bridge at various times. 
We scale the horizontal and vertical coordinates by $h_0$, which is presumably the only relevant length scale in the problem, and observe an excellent collapse of the data. 
This implies that the bridge growth exhibits self-similar dynamics, that we now set out to describe analytically.

\begin{figure*}
\includegraphics{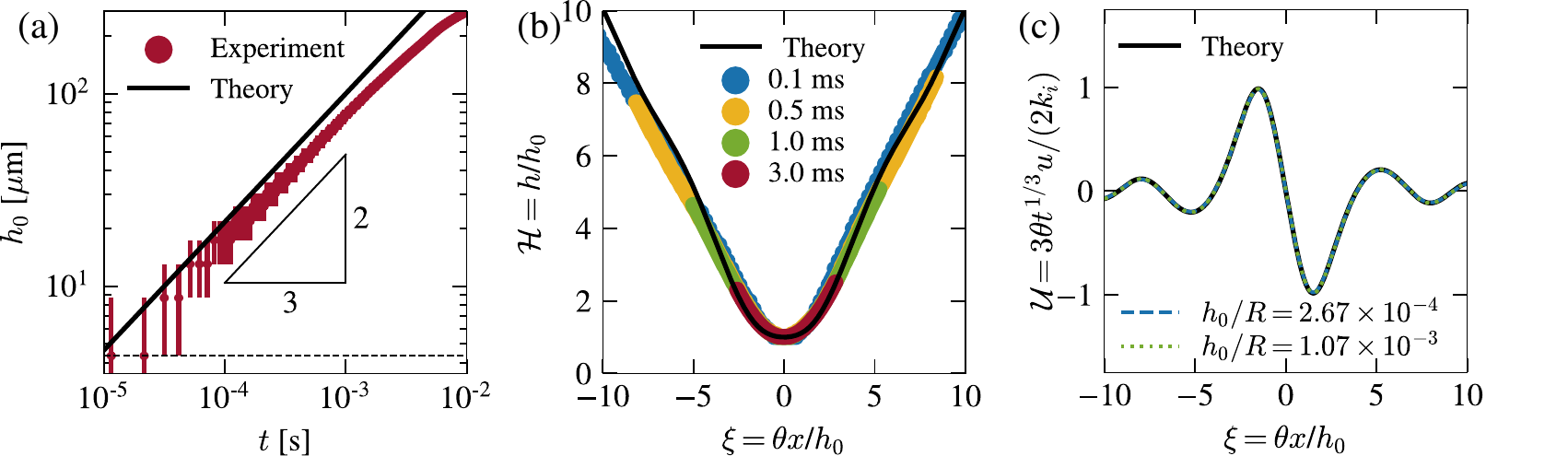}%
\caption{\label{fig3} Coalescence in the inertial regime. (a) Height of the bridge $h_0$ as a function of time after contact $t$ (dodecane lenses, $\theta = 29^\circ$, $\eta = 1.36$ mPa$\cdot$s, initial height $\approx~0.5$ mm). The solid line is the prediction from (\ref{eqK}). The horizontal dashed line indicates the resolution limit. (b) Rescaled experimental profiles at different times, $\mathcal{H} = h(x,t)/h_0(t)$ versus $\xi = x\theta/h_0(t)$. The collapse of the profiles indicates self-similar dynamics. The solid line is the similarity solution obtained from (\ref{eqMassDimensionless}, \ref{eqMomentumDimensionlessInertial}). (c) Rescaled velocity profile. The solid line is the similarity solution. The colored lines are numerical simulations for different values of $h_0/R$.}
\end{figure*}
%

\emph{Viscous and inertial similarity solutions.}---The main assumptions of our analysis are that (i) the flow during the initial stage of coalescence is predominantly parallel to the $xz$-plane (rendering the problem two-dimensional, following e.g. \cite{Ristenpart2006, Narhe2008, HernandezSanchez2012}), and (ii) the limiting mechanism for coalescence is the flow inside the drops (i.e. negligible flow inside the sub-phase, which in all but one experiment is at least one order of magnitude less viscous than the drop). 
Then, we can make use of the slender geometry of the system and use the thin-sheet equations \cite{Erneux1993, Scheid2012, Eggers2015}, 
\begin{eqnarray}
h_t + (uh)_x &=& 0, \label{eqMass} \\
\rho\left( u_t + uu_x \right)&=& \gamma h_{xxx} + 4 \eta \frac{(u_xh)_x}{h}, \label{eqMomentum}
\end{eqnarray}
which represent mass conservation and momentum conservation, respectively. 
Here, $h(x,t)$ is the shape of the bridge (Fig.~\ref{fig1}a), $u(x,t)$ is the horizontal velocity of the liquid inside the lenses (which is a plug flow to leading order in the slender approximation). 
The shape of the lens is assumed to be top-down symmetric, with uncertainty owing to the weak differences in surface tensions estimated to be less than 10\% (see Supplementary Material). 
We therefore take $\gamma$ as the surface tension of the lenses with respect to the surrounding air. 
The effect of gravity is expected to be negligible because the bridge is initially much smaller than the capillary length $\lambda_c = \sqrt{\gamma/(\Delta \rho g)} = \mathcal{O}(1)$ mm, and therefore we exclude it from the analysis.
Encouraged by the experiments, we search for similarity solutions of the form
\begin{equation}
h(x,t) = kt^\alpha \mathcal{H}(\xi), ~~u(x,t) = \frac{\alpha k}{\theta} t^\beta \mathcal{U}(\xi), ~~\xi = \frac{\theta x}{kt^\alpha},\label{eqGeneralAnsatz}
\end{equation}
where $\mathcal H$ and $\mathcal U$ are the similarity functions for the bridge profile and flow velocity. 
The choice of $\xi$ ensures that $h(x,t) \simeq \theta x$ far away from the bridge, in order to match a static solution with a contact angle $\theta$. 

We first examine the viscous regime, by setting $\rho \approx 0$ in (\ref{eqMass}, \ref{eqMomentum}). 
Inserting (\ref{eqGeneralAnsatz}) then readily leads to $\alpha=1$ and $\beta=0$, and explains the linear growth observed in the experiment. 
The parameter $k = k_v = \mathrm{d} h_0 / \mathrm{d} t$ thus provides the dimensional bridge velocity, and will be computed below. Equations (\ref{eqMass}, \ref{eqMomentum}) further reduce to 
\begin{eqnarray}
\mathcal{H} - \xi \mathcal{H}' + (\mathcal{H}\mathcal{U})' &=&0, \label{eqMassDimensionless} \\
\mathcal{H}\mathcal{H}''' + K_v (\mathcal{U}'\mathcal{H})' &=&0, \label{eqMomentumDimensionlessViscous}
\end{eqnarray}
providing a fourth order system of ODEs, that contains a parameter
\begin{equation}
K_v = \frac{4\eta k_v}{\gamma \theta^2}, \label{eqV} \label{eqV}
\end{equation}
representing the dimensionless bridge velocity. 
Hence, the selection of a unique solution requires five boundary conditions. 
We consider symmetric solutions and normalise the bridge height to unity at $\xi=0$, so that 
\begin{equation}
\mathcal{H}(0)=1, \quad \quad \mathcal{H}'(0)=0 \quad \mathrm{and} \quad \mathcal{U}(0)=0. \label{eqBoundaryConditionsZero}
\end{equation}

At large scale, this solution should match an initially static drop. 
This implies that the leading order asymptotics for large $\xi$ of $\mathcal H,~\mathcal U$ must correspond to time-independent $h,~u$. 
For the bridge profile, this implies $\mathcal H'(\infty) \rightarrow 1$, where we have also used the matching to the contact angle $\theta$. 
The velocity to leading order is $\mathcal U \simeq C \log \xi$ as $\xi~\rightarrow~\infty$; recalling that $\xi  \sim x/t$, a static drop at $t = 0$ corresponds to $C= 0$, which provides the 5th boundary condition.
The resulting boundary value problem is solved numerically by a shooting method, resulting in $K_v = 2.210$. 

We find excellent agreement between the experimental data and the similarity solution. 
The solid line in Fig.~\ref{fig2}a corresponds to the velocity prediction (\ref{eqV}) without any adjustable parameters. 
It is of interest to compare this result to the merging of drops on a solid substrate: owing to the no-slip boundary condition on a solid, coalescence is much slower on solid substrates with coalescence velocity $\sim~\theta^4$ \cite{HernandezSanchez2012} instead of $\sim~\theta^2$ observed for lenses.
In Fig.~\ref{fig2}b we compare the rescaled bridge profiles to $\mathcal{H}(\xi)$, shown as the solid line, and also find quantitative agreement. 
Figure~\ref{fig2}c shows the self-similar velocity $\mathcal{U}(\xi)$. 
Since the velocity inside the drop cannot be extracted from our experiments, we numerically solve the time-dependent equations (\ref{eqMass}, \ref{eqMomentum}) with $\rho=0$ using a finite element method and compare the result to the similarity solution. 
Details of the numerical method are found in the Supplementary Material. 
The numerical data in Fig.~\ref{fig2}c indeed collapse, and converge to the predicted similarity profile as $h_0/R \rightarrow 0$. 


The same scheme is followed for the regime where inertia dominates over viscosity, with the results outlined in Fig.~\ref{fig3}.  
Once again, we insert (\ref{eqGeneralAnsatz}), with $k=k_i$, in (\ref{eqMass},~\ref{eqMomentum}) but now in the inviscid limit ($\eta = 0$). 
The exponents can then be computed as $\alpha = 2/3$, $\beta = -1/3$, in agreement with the experiment. 
The momentum balance (\ref{eqMomentum}) now gives 
\begin{equation}
2\mathcal{U}\mathcal{U}' - \mathcal{U} - 2\xi \mathcal{U}'-K_i^{-1}\mathcal{H}''' = 0, \label{eqMomentumDimensionlessInertial}
\end{equation}
with a dimensionless constant 
\begin{equation}
K_i = \frac{2}{9} \frac{\rho k_i^3}{\gamma \theta^4}. \label{eqK}
\end{equation}
Mass conservation is unchanged as compared to (\ref{eqMassDimensionless}), so that we again require five boundary conditions to close the problem. 
As in the viscous case, four conditions follow from (\ref{eqBoundaryConditionsZero}) and $\mathcal H'(\infty) \rightarrow 1$. 
The 5th boundary condition again comes from the large-$\xi$ asymptotics -- one finds $\mathcal U \simeq C \xi^{-1/2}$ as $\xi \rightarrow \infty$ \cite{Ting1990}. This gives $u \simeq C x^{-1/2}$ which for a static outer drop {at $t=0$ implies $C=0$. 
Numerically solving the boundary value problem then gives $K_i = 0.106$. 

In Fig.~\ref{fig3}a we compare (\ref{eqK}) to experimental data of the lowest viscosity and find excellent agreement, without adjustable parameters. 
The spatial structure of the bridge also follows the predicted collapse, shown in Fig.~\ref{fig3}b, and agrees with the computed form $\mathcal H(\xi)$ (solid line). 
The dimensionless velocity $\mathcal{U}(\xi)$ is again compared to numerical simulations of (\ref{eqMass}, \ref{eqMomentum}) with $\eta \approx 0$, confirming the validity of the analysis. 
Interestingly, the velocity exhibits oscillations (Fig.~\ref{fig3}c) due to coalescence-induced inertio-capillary waves \cite{Billingham2005, Eddi2013}. 
These oscillations can indeed be predicted from the (higher order) asymptotics of the similarity equations \cite{Ting1990}. 
Let us remark that we cannot directly compare these results to the inertial coalescence on solid substrates, since the equivalent lubrication theory is not available owing to the no-slip condition.

\emph{Crossover.}---Several coalescence events in Fig.~\ref{fig1}b do not fit perfectly in either the viscous or in the inertial regime. 
As a final step we therefore describe the crossover between these regimes and collapse the entire set of experimental data. 
An estimate of the crossover height $h_\mathrm{c}$ and crossover time $t_\mathrm{c}$ can be obtained by setting $h_\mathrm{c} = k_v t_\mathrm{c} = k_i t_\mathrm{c}^{2/3}$, from which we find 
\begin{equation}
h_\mathrm{c} = \frac{k_i^3}{k_v^2} = \frac{72 K_i}{K_v^2}\frac{\eta^2}{\rho \gamma}, \quad t_\mathrm{c} = \frac{k_i^3}{k_v^3} = \frac{288 K_i}{K_v^3} \frac{\eta^3}{\rho \gamma^2 \theta^2}. \label{eqCrossoverScaling}
\end{equation}
Note that these are proportional to the intrinsic viscous scales $l_v = \eta^2/(\rho \gamma)$ and $t_v = \eta^3/(\rho \gamma^2)$ \cite{Eggers2008}, known for drop pinch-off, but with prefactors coming from the similarity analysis. 
Contrarily to pinch-off, however, we remark that the ultimate early-time coalescence is purely viscous \cite{Eggers1993}. 

\begin{figure}[b]
\includegraphics{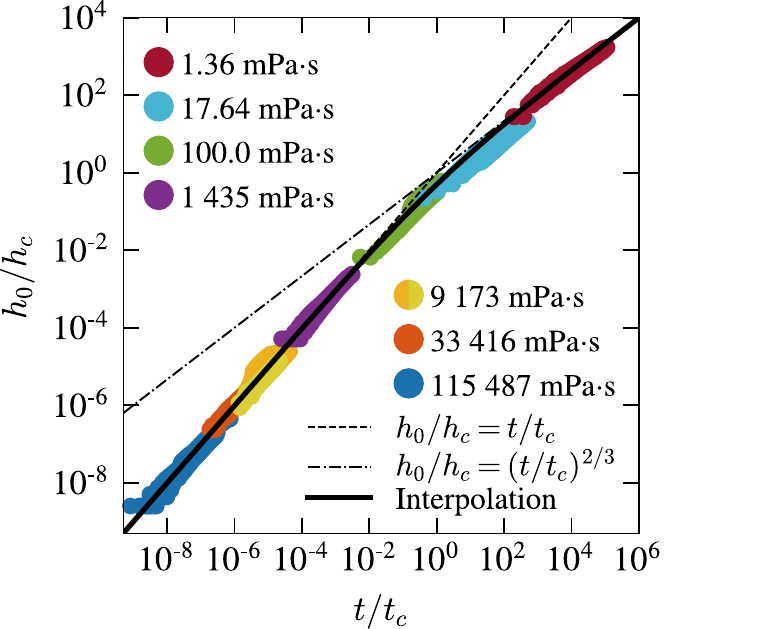}%
\caption{\label{fig4} Crossover between the viscous and inertial regimes, shown by a collapse of all experimental data on a master curve. Dashed line: viscous theory. Dotted-dashed line: inertial theory. Solid line: interpolation based on (\ref{eqInterpolation}). The lime-colored datapoints are with larger lens size ($R\approx4.1$~mm, compared to $R\approx2.5$~mm for all other data, showing that the dynamics do not depend on the drop size.}
\end{figure}

Figure~\ref{fig4} shows coalescence events for different viscosities (varied over five orders of magnitude), made dimensionless according to the crossover scales~(\ref{eqCrossoverScaling}). 
It is clear that the proposed scaling indeed collapses the data onto a single master curve, transitioning from the viscous to the inertial regime. 
In the spirit of the work on spherical drops~\cite{Paulsen2011,Xia2019} and drop impact \cite{Laan2014}, we propose an empirical formula based on a Pad\'e approximant which describes the two asymptotic regimes as well as the crossover region, 
\begin{equation}
h_0/h_\mathrm{c} = \left( \frac{1}{t/t_\mathrm{c}} + \frac{1}{(t/t_\mathrm{c})^{2/3}}\right)^{-1}. \label{eqInterpolation}
\end{equation}
We stress that, unlike the spherical drop case, the present interpolation (\ref{eqInterpolation}) contains no free parameters since $h_c$ and $t_c$ derived in (\ref{eqCrossoverScaling}) follow from the similarity solutions. 
The interpolation is superimposed as the solid line in Fig.~\ref{fig4}, providing an accurate description for all experiments.

\emph{Conclusion.}---Our results show that the coalescence of liquid lenses is accurately described by self-similar solutions to the thin-sheet equations. 
We have identified the crossover between the viscous regime and the inertial regime both experimentally and analytically. 
These coalescence dynamics are naturally very different from axisymmetric, spherical drops, though previous top-view experiments on liquid lenses \emph{did} observe axisymmetric-like dynamics \cite{Burton2007} -- the relation between horizontal and vertical growth remains to be understood. 
Importantly, the effect of the sub-phase viscosity is not included in our model -- and apparently it plays a subdominant role for the coalescence \cite{Paulsen2014}. 
Future work should be dedicated to more extreme cases, such as those where the viscosity of the sub-phase is much larger or where the layer thickness becomes small. 
This would be along the lines followed for the coalescence of circular nematic films \cite{Delabre2010}, where the influence of dissipation in the viscous sub-phase was systematically investigated. 
The present results provide a framework for such explorations, in particular the quantitative success of the thin-sheet equations, which will be of key interest to applications involving pre-wetted substrates. 

\begin{acknowledgments}
The authors thank Simon Hartmann, Sander Huisman and Herman Wijshoff for stimulating discussions.
We acknowledge support from an Industrial Partnership Programme of the Netherlands Organisation for Scientific Research (NWO), co-financed by Oc{\'e}-Technologies B.V., University of Twente, and Eindhoven University of Technology.  
Further support from the European Union's Horizon 2020 research and innovation programme LubISS (No. 722497), NWO Vici (No. 680-47-63) and the German Research Foundation (DFG-SPP 2171, HA8476/1, HA8467/2-1) is acknowledged. 
\end{acknowledgments}

\nocite{*}

\providecommand{\noopsort}[1]{}\providecommand{\singleletter}[1]{#1}%


\begin{thebibliography}{41}%
\makeatletter
\providecommand \@ifxundefined [1]{%
 \@ifx{#1\undefined}
}%
\providecommand \@ifnum [1]{%
 \ifnum #1\expandafter \@firstoftwo
 \else \expandafter \@secondoftwo
 \fi
}%
\providecommand \@ifx [1]{%
 \ifx #1\expandafter \@firstoftwo
 \else \expandafter \@secondoftwo
 \fi
}%
\providecommand \natexlab [1]{#1}%
\providecommand \enquote  [1]{``#1''}%
\providecommand \bibnamefont  [1]{#1}%
\providecommand \bibfnamefont [1]{#1}%
\providecommand \citenamefont [1]{#1}%
\providecommand \href@noop [0]{\@secondoftwo}%
\providecommand \href [0]{\begingroup \@sanitize@url \@href}%
\providecommand \@href[1]{\@@startlink{#1}\@@href}%
\providecommand \@@href[1]{\endgroup#1\@@endlink}%
\providecommand \@sanitize@url [0]{\catcode `\\12\catcode `\$12\catcode
  `\&12\catcode `\#12\catcode `\^12\catcode `\_12\catcode `\%12\relax}%
\providecommand \@@startlink[1]{}%
\providecommand \@@endlink[0]{}%
\providecommand \url  [0]{\begingroup\@sanitize@url \@url }%
\providecommand \@url [1]{\endgroup\@href {#1}{\urlprefix }}%
\providecommand \urlprefix  [0]{URL }%
\providecommand \Eprint [0]{\href }%
\providecommand \doibase [0]{https://doi.org/}%
\providecommand \selectlanguage [0]{\@gobble}%
\providecommand \bibinfo  [0]{\@secondoftwo}%
\providecommand \bibfield  [0]{\@secondoftwo}%
\providecommand \translation [1]{[#1]}%
\providecommand \BibitemOpen [0]{}%
\providecommand \bibitemStop [0]{}%
\providecommand \bibitemNoStop [0]{.\EOS\space}%
\providecommand \EOS [0]{\spacefactor3000\relax}%
\providecommand \BibitemShut  [1]{\csname bibitem#1\endcsname}%
\let\auto@bib@innerbib\@empty
\bibitem [{\citenamefont {Wijshoff}(2018)}]{Wijshoff2018}%
  \BibitemOpen
  \bibfield  {author} {\bibinfo {author} {\bibfnamefont {H.}~\bibnamefont
  {Wijshoff}},\ }\bibfield  {title} {\bibinfo {title} {Drop dynamics in the
  inkjet printing process},\ }\href
  {https://doi.org/10.1016/j.cocis.2017.11.004} {\bibfield  {journal} {\bibinfo
   {journal} {Curr. Opin. Colloid Interface Sci.}\ }\textbf {\bibinfo {volume}
  {36}},\ \bibinfo {pages} {20} (\bibinfo {year} {2018})}\BibitemShut {NoStop}%
\bibitem [{\citenamefont {Winkels}\ \emph {et~al.}(2011)\citenamefont
  {Winkels}, \citenamefont {Peters}, \citenamefont {Evangelista}, \citenamefont
  {Riepen}, \citenamefont {Daerr}, \citenamefont {Limat},\ and\ \citenamefont
  {Snoeijer}}]{Winkels2011}%
  \BibitemOpen
  \bibfield  {author} {\bibinfo {author} {\bibfnamefont {K.~G.}\ \bibnamefont
  {Winkels}}, \bibinfo {author} {\bibfnamefont {I.~R.}\ \bibnamefont {Peters}},
  \bibinfo {author} {\bibfnamefont {F.}~\bibnamefont {Evangelista}}, \bibinfo
  {author} {\bibfnamefont {M.}~\bibnamefont {Riepen}}, \bibinfo {author}
  {\bibfnamefont {A.}~\bibnamefont {Daerr}}, \bibinfo {author} {\bibfnamefont
  {L.}~\bibnamefont {Limat}},\ and\ \bibinfo {author} {\bibfnamefont {J.~H.}\
  \bibnamefont {Snoeijer}},\ }\bibfield  {title} {\bibinfo {title} {Receding
  contact lines: From sliding drops to immersion lithography},\ }\href
  {https://doi.org/10.1140/epjst/e2011-01374-6} {\bibfield  {journal} {\bibinfo
   {journal} {Eur. Phys. J. Special Topics}\ }\textbf {\bibinfo {volume}
  {192}},\ \bibinfo {pages} {195} (\bibinfo {year} {2011})}\BibitemShut
  {NoStop}%
\bibitem [{\citenamefont {Grabowski}\ and\ \citenamefont
  {Wang}(2013)}]{Grabowski2013}%
  \BibitemOpen
  \bibfield  {author} {\bibinfo {author} {\bibfnamefont {W.~W.}\ \bibnamefont
  {Grabowski}}\ and\ \bibinfo {author} {\bibfnamefont {L.-P.}\ \bibnamefont
  {Wang}},\ }\bibfield  {title} {\bibinfo {title} {Growth of cloud droplets in
  turbulent environment},\ }\href
  {https://doi.org/10.1146/annurev-fluid-011212-140750} {\bibfield  {journal}
  {\bibinfo  {journal} {Annu.~Rev. Fluid~Mech.}\ }\textbf {\bibinfo {volume}
  {45}},\ \bibinfo {pages} {293} (\bibinfo {year} {2013})}\BibitemShut
  {NoStop}%
\bibitem [{\citenamefont {Wisdom}\ \emph {et~al.}(2013)\citenamefont {Wisdom},
  \citenamefont {Watson}, \citenamefont {Qu}, \citenamefont {Liu},
  \citenamefont {Watson},\ and\ \citenamefont {Chen}}]{Wisdom2013}%
  \BibitemOpen
  \bibfield  {author} {\bibinfo {author} {\bibfnamefont {K.~M.}\ \bibnamefont
  {Wisdom}}, \bibinfo {author} {\bibfnamefont {J.~A.}\ \bibnamefont {Watson}},
  \bibinfo {author} {\bibfnamefont {X.}~\bibnamefont {Qu}}, \bibinfo {author}
  {\bibfnamefont {F.}~\bibnamefont {Liu}}, \bibinfo {author} {\bibfnamefont
  {G.~S.}\ \bibnamefont {Watson}},\ and\ \bibinfo {author} {\bibfnamefont
  {C.-H.}\ \bibnamefont {Chen}},\ }\bibfield  {title} {\bibinfo {title}
  {Self-cleaning of superhydrophobic surfaces by self-propelled jumping
  condensate},\ }\href {https://doi.org/10.1073/pnas.1210770110} {\bibfield
  {journal} {\bibinfo  {journal} {Proc. Natl. Acad. Sci. U.S.A.}\ }\textbf
  {\bibinfo {volume} {20}},\ \bibinfo {pages} {7992} (\bibinfo {year}
  {2013})}\BibitemShut {NoStop}%
\bibitem [{\citenamefont {Watson}\ \emph {et~al.}(2014)\citenamefont {Watson},
  \citenamefont {Gellender},\ and\ \citenamefont {Watson}}]{Watson2014}%
  \BibitemOpen
  \bibfield  {author} {\bibinfo {author} {\bibfnamefont {G.~S.}\ \bibnamefont
  {Watson}}, \bibinfo {author} {\bibfnamefont {M.}~\bibnamefont {Gellender}},\
  and\ \bibinfo {author} {\bibfnamefont {J.~A.}\ \bibnamefont {Watson}},\
  }\bibfield  {title} {\bibinfo {title} {Self-propulsion of dew drops on lotus
  leaves: a potential mechanism for self-cleaning},\ }\href
  {https://doi.org/10.1080/08927014.2014.880885} {\bibfield  {journal}
  {\bibinfo  {journal} {Biofouling}\ }\textbf {\bibinfo {volume} {30}},\
  \bibinfo {pages} {427} (\bibinfo {year} {2014})}\BibitemShut {NoStop}%
\bibitem [{\citenamefont {Eggers}\ \emph {et~al.}(1999)\citenamefont {Eggers},
  \citenamefont {Lister},\ and\ \citenamefont {Stone}}]{Eggers1999}%
  \BibitemOpen
  \bibfield  {author} {\bibinfo {author} {\bibfnamefont {J.}~\bibnamefont
  {Eggers}}, \bibinfo {author} {\bibfnamefont {J.~R.}\ \bibnamefont {Lister}},\
  and\ \bibinfo {author} {\bibfnamefont {H.~A.}\ \bibnamefont {Stone}},\
  }\bibfield  {title} {\bibinfo {title} {Coalescence of liquid drops},\ }\href
  {https://doi.org/10.1017/S002211209900662X} {\bibfield  {journal} {\bibinfo
  {journal} {J. Fluid Mech.}\ }\textbf {\bibinfo {volume} {401}},\ \bibinfo
  {pages} {293} (\bibinfo {year} {1999})}\BibitemShut {NoStop}%
\bibitem [{\citenamefont {Duchemin}\ \emph {et~al.}(2003)\citenamefont
  {Duchemin}, \citenamefont {Eggers},\ and\ \citenamefont
  {Josserand}}]{Duchemin2003}%
  \BibitemOpen
  \bibfield  {author} {\bibinfo {author} {\bibfnamefont {L.}~\bibnamefont
  {Duchemin}}, \bibinfo {author} {\bibfnamefont {J.}~\bibnamefont {Eggers}},\
  and\ \bibinfo {author} {\bibfnamefont {C.}~\bibnamefont {Josserand}},\
  }\bibfield  {title} {\bibinfo {title} {Inviscid coalescence of drops},\
  }\href {https://doi.org/10.1017/S0022112003004646} {\bibfield  {journal}
  {\bibinfo  {journal} {J. Fluid Mech.}\ }\textbf {\bibinfo {volume} {487}},\
  \bibinfo {pages} {167} (\bibinfo {year} {2003})}\BibitemShut {NoStop}%
\bibitem [{\citenamefont {Aarts}\ \emph {et~al.}(2005)\citenamefont {Aarts},
  \citenamefont {Lekkerkerker}, \citenamefont {Guo}, \citenamefont {Wegdam},\
  and\ \citenamefont {Bonn}}]{Aarts2005}%
  \BibitemOpen
  \bibfield  {author} {\bibinfo {author} {\bibfnamefont {D.~G. A.~L.}\
  \bibnamefont {Aarts}}, \bibinfo {author} {\bibfnamefont {H.~N.~W.}\
  \bibnamefont {Lekkerkerker}}, \bibinfo {author} {\bibfnamefont
  {H.}~\bibnamefont {Guo}}, \bibinfo {author} {\bibfnamefont {G.~H.}\
  \bibnamefont {Wegdam}},\ and\ \bibinfo {author} {\bibfnamefont
  {D.}~\bibnamefont {Bonn}},\ }\bibfield  {title} {\bibinfo {title}
  {Hydrodynamics of droplet coalescence},\ }\href
  {https://doi.org/10.1103/PhysRevLett.95.164503} {\bibfield  {journal}
  {\bibinfo  {journal} {Phys. Rev. Lett.}\ }\textbf {\bibinfo {volume} {95}},\
  \bibinfo {pages} {164503} (\bibinfo {year} {2005})}\BibitemShut {NoStop}%
\bibitem [{\citenamefont {Thoroddsen}\ \emph {et~al.}(2007)\citenamefont
  {Thoroddsen}, \citenamefont {Qian}, \citenamefont {Etoh},\ and\ \citenamefont
  {Takehara}}]{Thoroddsen2007}%
  \BibitemOpen
  \bibfield  {author} {\bibinfo {author} {\bibfnamefont {S.~T.}\ \bibnamefont
  {Thoroddsen}}, \bibinfo {author} {\bibfnamefont {B.}~\bibnamefont {Qian}},
  \bibinfo {author} {\bibfnamefont {T.~G.}\ \bibnamefont {Etoh}},\ and\
  \bibinfo {author} {\bibfnamefont {K.}~\bibnamefont {Takehara}},\ }\bibfield
  {title} {\bibinfo {title} {The initial coalescence of miscible drops},\
  }\href {https://doi.org/10.1063/1.2746382} {\bibfield  {journal} {\bibinfo
  {journal} {Phys. Fluids}\ }\textbf {\bibinfo {volume} {19}},\ \bibinfo
  {pages} {072110} (\bibinfo {year} {2007})}\BibitemShut {NoStop}%
\bibitem [{\citenamefont {Paulsen}\ \emph {et~al.}(2011)\citenamefont
  {Paulsen}, \citenamefont {Burton},\ and\ \citenamefont
  {Nagel}}]{Paulsen2011}%
  \BibitemOpen
  \bibfield  {author} {\bibinfo {author} {\bibfnamefont {J.~D.}\ \bibnamefont
  {Paulsen}}, \bibinfo {author} {\bibfnamefont {J.~C.}\ \bibnamefont
  {Burton}},\ and\ \bibinfo {author} {\bibfnamefont {S.~R.}\ \bibnamefont
  {Nagel}},\ }\bibfield  {title} {\bibinfo {title} {Viscous to inertial
  crossover in liquid drop coalescence},\ }\href
  {https://doi.org/10.1103/PhysRevLett.106.114501} {\bibfield  {journal}
  {\bibinfo  {journal} {Phys. Rev. Lett.}\ }\textbf {\bibinfo {volume} {106}},\
  \bibinfo {pages} {114501} (\bibinfo {year} {2011})}\BibitemShut {NoStop}%
\bibitem [{\citenamefont {Ristenpart}\ \emph {et~al.}(2006)\citenamefont
  {Ristenpart}, \citenamefont {McCalla}, \citenamefont {Roy},\ and\
  \citenamefont {Stone}}]{Ristenpart2006}%
  \BibitemOpen
  \bibfield  {author} {\bibinfo {author} {\bibfnamefont {W.~D.}\ \bibnamefont
  {Ristenpart}}, \bibinfo {author} {\bibfnamefont {P.~M.}\ \bibnamefont
  {McCalla}}, \bibinfo {author} {\bibfnamefont {R.~V.}\ \bibnamefont {Roy}},\
  and\ \bibinfo {author} {\bibfnamefont {H.~A.}\ \bibnamefont {Stone}},\
  }\bibfield  {title} {\bibinfo {title} {Coalescence of spreading droplets on a
  wettable substrate},\ }\href {https://doi.org/10.1103/PhysRevLett.97.064501}
  {\bibfield  {journal} {\bibinfo  {journal} {Phys. Rev. Lett.}\ }\textbf
  {\bibinfo {volume} {97}},\ \bibinfo {pages} {064501} (\bibinfo {year}
  {2006})}\BibitemShut {NoStop}%
\bibitem [{\citenamefont {Narge}\ \emph {et~al.}(2008)\citenamefont {Narge},
  \citenamefont {Beysens},\ and\ \citenamefont {Pomeau}}]{Narhe2008}%
  \BibitemOpen
  \bibfield  {author} {\bibinfo {author} {\bibfnamefont {R.~D.}\ \bibnamefont
  {Narge}}, \bibinfo {author} {\bibfnamefont {D.~A.}\ \bibnamefont {Beysens}},\
  and\ \bibinfo {author} {\bibfnamefont {Y.}~\bibnamefont {Pomeau}},\
  }\bibfield  {title} {\bibinfo {title} {Dynamics drying in the early-stage
  coalescence of droplets sitting on a plate},\ }\href
  {https://doi.org/10.1209/0295-5075/81/46002} {\bibfield  {journal} {\bibinfo
  {journal} {Eur. Phys. Lett.}\ }\textbf {\bibinfo {volume} {81}},\ \bibinfo
  {pages} {46002} (\bibinfo {year} {2008})}\BibitemShut {NoStop}%
\bibitem [{\citenamefont {Lee}\ \emph {et~al.}(2012)\citenamefont {Lee},
  \citenamefont {Kang}, \citenamefont {Yoon},\ and\ \citenamefont
  {Yarin}}]{Lee2012}%
  \BibitemOpen
  \bibfield  {author} {\bibinfo {author} {\bibfnamefont {M.~W.}\ \bibnamefont
  {Lee}}, \bibinfo {author} {\bibfnamefont {D.~K.}\ \bibnamefont {Kang}},
  \bibinfo {author} {\bibfnamefont {S.~S.}\ \bibnamefont {Yoon}},\ and\
  \bibinfo {author} {\bibfnamefont {A.~L.}\ \bibnamefont {Yarin}},\ }\bibfield
  {title} {\bibinfo {title} {Coalescence of two drops on partially wettable
  substrates},\ }\href {https://doi.org/10.1021/la204867c} {\bibfield
  {journal} {\bibinfo  {journal} {Langmuir}\ }\textbf {\bibinfo {volume}
  {28}},\ \bibinfo {pages} {3791} (\bibinfo {year} {2012})}\BibitemShut
  {NoStop}%
\bibitem [{\citenamefont {Hern\'andez-S\'anchez}\ \emph
  {et~al.}(2012)\citenamefont {Hern\'andez-S\'anchez}, \citenamefont {Lubbers},
  \citenamefont {Eddi},\ and\ \citenamefont {Snoeijer}}]{HernandezSanchez2012}%
  \BibitemOpen
  \bibfield  {author} {\bibinfo {author} {\bibfnamefont {J.~F.}\ \bibnamefont
  {Hern\'andez-S\'anchez}}, \bibinfo {author} {\bibfnamefont {L.~A.}\
  \bibnamefont {Lubbers}}, \bibinfo {author} {\bibfnamefont {A.}~\bibnamefont
  {Eddi}},\ and\ \bibinfo {author} {\bibfnamefont {J.~H.}\ \bibnamefont
  {Snoeijer}},\ }\bibfield  {title} {\bibinfo {title} {Symmetric and asymmetric
  coalescence of drops on a substrate},\ }\href
  {https://doi.org/10.1103/PhysRevLett.109.184502} {\bibfield  {journal}
  {\bibinfo  {journal} {Phys.~Rev.~Lett.}\ }\textbf {\bibinfo {volume} {109}},\
  \bibinfo {pages} {184502} (\bibinfo {year} {2012})}\BibitemShut {NoStop}%
\bibitem [{\citenamefont {Eddi}\ \emph {et~al.}(2013)\citenamefont {Eddi},
  \citenamefont {Winkels},\ and\ \citenamefont {Snoeijer}}]{Eddi2013}%
  \BibitemOpen
  \bibfield  {author} {\bibinfo {author} {\bibfnamefont {A.}~\bibnamefont
  {Eddi}}, \bibinfo {author} {\bibfnamefont {K.~G.}\ \bibnamefont {Winkels}},\
  and\ \bibinfo {author} {\bibfnamefont {J.~H.}\ \bibnamefont {Snoeijer}},\
  }\bibfield  {title} {\bibinfo {title} {Influence of droplet geometry on the
  coalescence of low viscosity drops},\ }\href
  {https://doi.org/10.1103/PhysRevLett.111.144502} {\bibfield  {journal}
  {\bibinfo  {journal} {Phys.~Rev.~Lett.}\ }\textbf {\bibinfo {volume} {111}},\
  \bibinfo {pages} {144502} (\bibinfo {year} {2013})}\BibitemShut {NoStop}%
\bibitem [{\citenamefont {Burton}\ and\ \citenamefont
  {Taborek}(2007)}]{Burton2007}%
  \BibitemOpen
  \bibfield  {author} {\bibinfo {author} {\bibfnamefont {J.~C.}\ \bibnamefont
  {Burton}}\ and\ \bibinfo {author} {\bibfnamefont {P.}~\bibnamefont
  {Taborek}},\ }\bibfield  {title} {\bibinfo {title} {Role of dimensionality
  and axisymmetry in fluid pinch-off and coalescence},\ }\href
  {https://doi.org/10.1103/PhysRevLett.98.224502} {\bibfield  {journal}
  {\bibinfo  {journal} {Phys. Rev. Lett.}\ }\textbf {\bibinfo {volume} {98}},\
  \bibinfo {pages} {224502} (\bibinfo {year} {2007})}\BibitemShut {NoStop}%
\bibitem [{\citenamefont {Delabre}\ and\ \citenamefont
  {Cazabat}(2010)}]{Delabre2010}%
  \BibitemOpen
  \bibfield  {author} {\bibinfo {author} {\bibfnamefont {U.}~\bibnamefont
  {Delabre}}\ and\ \bibinfo {author} {\bibfnamefont {A.-M.}\ \bibnamefont
  {Cazabat}},\ }\bibfield  {title} {\bibinfo {title} {Coalescence driven by
  line tension in thin nematic films},\ }\href
  {https://doi.org/10.1103/PhysRevLett.104.227801} {\bibfield  {journal}
  {\bibinfo  {journal} {Phys. Rev. Lett.}\ }\textbf {\bibinfo {volume} {104}},\
  \bibinfo {pages} {227801} (\bibinfo {year} {2010})}\BibitemShut {NoStop}%
\bibitem [{\citenamefont {Sokuler}\ \emph {et~al.}(2009)\citenamefont
  {Sokuler}, \citenamefont {Auernhammer}, \citenamefont {Roth}, \citenamefont
  {Liu}, \citenamefont {Bonacurrso},\ and\ \citenamefont
  {B\"utt}}]{Sokuler2009}%
  \BibitemOpen
  \bibfield  {author} {\bibinfo {author} {\bibfnamefont {M.}~\bibnamefont
  {Sokuler}}, \bibinfo {author} {\bibfnamefont {G.~K.}\ \bibnamefont
  {Auernhammer}}, \bibinfo {author} {\bibfnamefont {M.}~\bibnamefont {Roth}},
  \bibinfo {author} {\bibfnamefont {C.}~\bibnamefont {Liu}}, \bibinfo {author}
  {\bibfnamefont {E.}~\bibnamefont {Bonacurrso}},\ and\ \bibinfo {author}
  {\bibfnamefont {H.-J.}\ \bibnamefont {B\"utt}},\ }\bibfield  {title}
  {\bibinfo {title} {The softer the better: Fast condensation on soft
  surfaces},\ }\href {https://doi.org/10.1021/la903996j} {\bibfield  {journal}
  {\bibinfo  {journal} {Langmuir}\ }\textbf {\bibinfo {volume} {26}},\ \bibinfo
  {pages} {1544} (\bibinfo {year} {2009})}\BibitemShut {NoStop}%
\bibitem [{\citenamefont {Anand}\ \emph {et~al.}(2012)\citenamefont {Anand},
  \citenamefont {Paxson}, \citenamefont {Dhiman}, \citenamefont {Smith},\ and\
  \citenamefont {Varanasi}}]{Anand2012}%
  \BibitemOpen
  \bibfield  {author} {\bibinfo {author} {\bibfnamefont {S.}~\bibnamefont
  {Anand}}, \bibinfo {author} {\bibfnamefont {A.~T.}\ \bibnamefont {Paxson}},
  \bibinfo {author} {\bibfnamefont {R.}~\bibnamefont {Dhiman}}, \bibinfo
  {author} {\bibfnamefont {J.~D.}\ \bibnamefont {Smith}},\ and\ \bibinfo
  {author} {\bibfnamefont {K.~K.}\ \bibnamefont {Varanasi}},\ }\bibfield
  {title} {\bibinfo {title} {Enhanced condensation on lubricant-impregnated
  nanotextured surfaces},\ }\href {https://doi.org/10.1021/nn303867y}
  {\bibfield  {journal} {\bibinfo  {journal} {ACS Nano}\ }\textbf {\bibinfo
  {volume} {6}},\ \bibinfo {pages} {10122} (\bibinfo {year}
  {2012})}\BibitemShut {NoStop}%
\bibitem [{\citenamefont {Kim}\ \emph {et~al.}(2012)\citenamefont {Kim},
  \citenamefont {Wong}, \citenamefont {Alvarenga}, \citenamefont {Kreder},
  \citenamefont {Adorno-Martinez},\ and\ \citenamefont {Aizenberg}}]{Kim2012}%
  \BibitemOpen
  \bibfield  {author} {\bibinfo {author} {\bibfnamefont {P.}~\bibnamefont
  {Kim}}, \bibinfo {author} {\bibfnamefont {T.-S.}\ \bibnamefont {Wong}},
  \bibinfo {author} {\bibfnamefont {J.}~\bibnamefont {Alvarenga}}, \bibinfo
  {author} {\bibfnamefont {M.~J.}\ \bibnamefont {Kreder}}, \bibinfo {author}
  {\bibfnamefont {W.~E.}\ \bibnamefont {Adorno-Martinez}},\ and\ \bibinfo
  {author} {\bibfnamefont {J.}~\bibnamefont {Aizenberg}},\ }\bibfield  {title}
  {\bibinfo {title} {Liquid-infused nanostructured surfaces with extreme
  anti-ice and anti-frost performance},\ }\href
  {https://doi.org/10.1021/nn302310q} {\bibfield  {journal} {\bibinfo
  {journal} {ACS Nano}\ }\textbf {\bibinfo {volume} {6}},\ \bibinfo {pages}
  {6569} (\bibinfo {year} {2012})}\BibitemShut {NoStop}%
\bibitem [{\citenamefont {Hack}\ \emph {et~al.}(2018)\citenamefont {Hack},
  \citenamefont {Costalonga}, \citenamefont {Segers}, \citenamefont
  {Karpitschka}, \citenamefont {Wijshoff},\ and\ \citenamefont
  {Snoeijer}}]{Hack2018}%
  \BibitemOpen
  \bibfield  {author} {\bibinfo {author} {\bibfnamefont {M.~A.}\ \bibnamefont
  {Hack}}, \bibinfo {author} {\bibfnamefont {M.}~\bibnamefont {Costalonga}},
  \bibinfo {author} {\bibfnamefont {T.}~\bibnamefont {Segers}}, \bibinfo
  {author} {\bibfnamefont {S.}~\bibnamefont {Karpitschka}}, \bibinfo {author}
  {\bibfnamefont {H.}~\bibnamefont {Wijshoff}},\ and\ \bibinfo {author}
  {\bibfnamefont {J.~H.}\ \bibnamefont {Snoeijer}},\ }\bibfield  {title}
  {\bibinfo {title} {Printing wet-on-wet: Attraction and repulsion of drops on
  a viscous film},\ }\href {https://doi.org/10.1063/1.5048681} {\bibfield
  {journal} {\bibinfo  {journal} {Appl. Phys. Lett.}\ }\textbf {\bibinfo
  {volume} {113}},\ \bibinfo {pages} {183701} (\bibinfo {year}
  {2018})}\BibitemShut {NoStop}%
\bibitem [{\citenamefont {Wasan}\ \emph {et~al.}(1978)\citenamefont {Wasan},
  \citenamefont {Shah}, \citenamefont {Aderangi}, \citenamefont {Chan},\ and\
  \citenamefont {McNamara}}]{Wasan1978}%
  \BibitemOpen
  \bibfield  {author} {\bibinfo {author} {\bibfnamefont {D.~T.}\ \bibnamefont
  {Wasan}}, \bibinfo {author} {\bibfnamefont {S.~M.}\ \bibnamefont {Shah}},
  \bibinfo {author} {\bibfnamefont {N.}~\bibnamefont {Aderangi}}, \bibinfo
  {author} {\bibfnamefont {M.~S.}\ \bibnamefont {Chan}},\ and\ \bibinfo
  {author} {\bibfnamefont {J.~J.}\ \bibnamefont {McNamara}},\ }\bibfield
  {title} {\bibinfo {title} {Observations on the coalescence behavior of oil
  droplets and emulsion stability in enhanced oil recovery},\ }\href
  {https://doi.org/10.2118/6846-PA} {\bibfield  {journal} {\bibinfo  {journal}
  {SPE J.}\ }\textbf {\bibinfo {volume} {18}},\ \bibinfo {pages} {409}
  (\bibinfo {year} {1978})}\BibitemShut {NoStop}%
\bibitem [{\citenamefont {Wasan}\ \emph {et~al.}(1979)\citenamefont {Wasan},
  \citenamefont {McNamara}, \citenamefont {Shah}, \citenamefont {Sampath},\
  and\ \citenamefont {Aderangi}}]{Wasan1979}%
  \BibitemOpen
  \bibfield  {author} {\bibinfo {author} {\bibfnamefont {D.~T.}\ \bibnamefont
  {Wasan}}, \bibinfo {author} {\bibfnamefont {J.~J.}\ \bibnamefont {McNamara}},
  \bibinfo {author} {\bibfnamefont {S.~M.}\ \bibnamefont {Shah}}, \bibinfo
  {author} {\bibfnamefont {K.}~\bibnamefont {Sampath}},\ and\ \bibinfo {author}
  {\bibfnamefont {N.}~\bibnamefont {Aderangi}},\ }\bibfield  {title} {\bibinfo
  {title} {The role of coalescence phenomena and interfacial rheological
  properties in enhanced oil recovery: An overview},\ }\href
  {https://doi.org/10.1122/1.549524} {\bibfield  {journal} {\bibinfo  {journal}
  {J. Rheol.}\ }\textbf {\bibinfo {volume} {23}},\ \bibinfo {pages} {181}
  (\bibinfo {year} {1979})}\BibitemShut {NoStop}%
\bibitem [{\citenamefont {Shaw}(2003)}]{Shaw2003}%
  \BibitemOpen
  \bibfield  {author} {\bibinfo {author} {\bibfnamefont {J.~M.}\ \bibnamefont
  {Shaw}},\ }\bibfield  {title} {\bibinfo {title} {A microscopic view of oil
  slick break-up and emulsion formation in breaking waves},\ }\href
  {https://doi.org/10.1016/S1353-2561(03)00061-6} {\bibfield  {journal}
  {\bibinfo  {journal} {Spill Sci. Technol. Bull.}\ }\textbf {\bibinfo {volume}
  {8}},\ \bibinfo {pages} {491} (\bibinfo {year} {2003})}\BibitemShut {NoStop}%
\bibitem [{\citenamefont {Kamp}\ \emph {et~al.}(2016)\citenamefont {Kamp},
  \citenamefont {Villwock},\ and\ \citenamefont {Kraume}}]{Kamp2016}%
  \BibitemOpen
  \bibfield  {author} {\bibinfo {author} {\bibfnamefont {J.}~\bibnamefont
  {Kamp}}, \bibinfo {author} {\bibfnamefont {J.}~\bibnamefont {Villwock}},\
  and\ \bibinfo {author} {\bibfnamefont {M.}~\bibnamefont {Kraume}},\
  }\bibfield  {title} {\bibinfo {title} {Drop coalescence in technical
  liquid/liquid applications: a review on experimental techniques and modeling
  approaches},\ }\href {https://doi.org/10.1515/revce-2015-0071} {\bibfield
  {journal} {\bibinfo  {journal} {Rev. Chem. Eng.}\ }\textbf {\bibinfo {volume}
  {33}},\ \bibinfo {pages} {1} (\bibinfo {year} {2016})}\BibitemShut {NoStop}%
\bibitem [{\citenamefont {Keiser}\ \emph {et~al.}(2017)\citenamefont {Keiser},
  \citenamefont {Bense}, \citenamefont {Colinet}, \citenamefont {Bico},\ and\
  \citenamefont {Reyssat}}]{Keiser2017}%
  \BibitemOpen
  \bibfield  {author} {\bibinfo {author} {\bibfnamefont {L.}~\bibnamefont
  {Keiser}}, \bibinfo {author} {\bibfnamefont {H.}~\bibnamefont {Bense}},
  \bibinfo {author} {\bibfnamefont {P.}~\bibnamefont {Colinet}}, \bibinfo
  {author} {\bibfnamefont {J.}~\bibnamefont {Bico}},\ and\ \bibinfo {author}
  {\bibfnamefont {E.}~\bibnamefont {Reyssat}},\ }\bibfield  {title} {\bibinfo
  {title} {Marangoni bursting: Evaporation-induced emulsification of binary
  mixtures on a liquid layer},\ }\href
  {https://doi.org/10.1103/PhysRevLett.118.074504} {\bibfield  {journal}
  {\bibinfo  {journal} {Phys. Rev. Lett.}\ }\textbf {\bibinfo {volume} {118}},\
  \bibinfo {pages} {074504} (\bibinfo {year} {2017})}\BibitemShut {NoStop}%
\bibitem [{\citenamefont {Smith}\ \emph {et~al.}(2013)\citenamefont {Smith},
  \citenamefont {Dhiman}, \citenamefont {Anand}, \citenamefont {Reza-Garduno},
  \citenamefont {Cohen}, \citenamefont {McKinley},\ and\ \citenamefont
  {Varanasi}}]{Smith2013}%
  \BibitemOpen
  \bibfield  {author} {\bibinfo {author} {\bibfnamefont {J.~D.}\ \bibnamefont
  {Smith}}, \bibinfo {author} {\bibfnamefont {R.}~\bibnamefont {Dhiman}},
  \bibinfo {author} {\bibfnamefont {S.}~\bibnamefont {Anand}}, \bibinfo
  {author} {\bibfnamefont {E.}~\bibnamefont {Reza-Garduno}}, \bibinfo {author}
  {\bibfnamefont {R.~E.}\ \bibnamefont {Cohen}}, \bibinfo {author}
  {\bibfnamefont {G.~H.}\ \bibnamefont {McKinley}},\ and\ \bibinfo {author}
  {\bibfnamefont {K.~K.}\ \bibnamefont {Varanasi}},\ }\bibfield  {title}
  {\bibinfo {title} {Droplet mobility on lubricant-impregnated surfaces},\
  }\href {https://doi.org/10.1039/C2SM27032C} {\bibfield  {journal} {\bibinfo
  {journal} {Soft Matter}\ }\textbf {\bibinfo {volume} {9}},\ \bibinfo {pages}
  {1772} (\bibinfo {year} {2013})}\BibitemShut {NoStop}%
\bibitem [{\citenamefont {Langmuir}(1933)}]{Langmuir1933}%
  \BibitemOpen
  \bibfield  {author} {\bibinfo {author} {\bibfnamefont {I.}~\bibnamefont
  {Langmuir}},\ }\bibfield  {title} {\bibinfo {title} {Oil lenses on water and
  the nature of monomolecular expanded films},\ }\href
  {https://doi.org/10.1063/1.1749243} {\bibfield  {journal} {\bibinfo
  {journal} {J. Chem. Phys.}\ }\textbf {\bibinfo {volume} {1}},\ \bibinfo
  {pages} {756} (\bibinfo {year} {1933})}\BibitemShut {NoStop}%
\bibitem [{\citenamefont {de~Gennes}\ \emph {et~al.}(2004)\citenamefont
  {de~Gennes}, \citenamefont {Brochard-Wyart},\ and\ \citenamefont
  {Qu{\'e}r{\'e}}}]{deGennes2004}%
  \BibitemOpen
  \bibfield  {author} {\bibinfo {author} {\bibfnamefont {P.-G.}\ \bibnamefont
  {de~Gennes}}, \bibinfo {author} {\bibfnamefont {F.}~\bibnamefont
  {Brochard-Wyart}},\ and\ \bibinfo {author} {\bibfnamefont {D.}~\bibnamefont
  {Qu{\'e}r{\'e}}},\ }\href {https://doi.org/10.1007/978-0-387-21656-0} {\emph
  {\bibinfo {title} {Capillarity and Wetting Phenomena: Drops, Bubbles, Pearls,
  Waves}}}\ (\bibinfo  {publisher} {Springer},\ \bibinfo {year}
  {2004})\BibitemShut {NoStop}%
\bibitem [{\citenamefont {Xia}\ \emph {et~al.}(2019)\citenamefont {Xia},
  \citenamefont {He},\ and\ \citenamefont {Zhang}}]{Xia2019}%
  \BibitemOpen
  \bibfield  {author} {\bibinfo {author} {\bibfnamefont {X.}~\bibnamefont
  {Xia}}, \bibinfo {author} {\bibfnamefont {C.}~\bibnamefont {He}},\ and\
  \bibinfo {author} {\bibfnamefont {P.}~\bibnamefont {Zhang}},\ }\bibfield
  {title} {\bibinfo {title} {Universality in the viscous-to-inertial
  coalescence of liquid droplets},\ }\href
  {https://doi.org/10.1073/pnas.1910711116} {\bibfield  {journal} {\bibinfo
  {journal} {Proc. Natl. Acad. Sci. U.S.A.}\ }\textbf {\bibinfo {volume}
  {116}},\ \bibinfo {pages} {23467} (\bibinfo {year} {2019})}\BibitemShut
  {NoStop}%
\bibitem [{\citenamefont {Hansen}\ and\ \citenamefont
  {R{\o}dsrud}(1991)}]{Hansen1991}%
  \BibitemOpen
  \bibfield  {author} {\bibinfo {author} {\bibfnamefont {F.~K.}\ \bibnamefont
  {Hansen}}\ and\ \bibinfo {author} {\bibfnamefont {G.}~\bibnamefont
  {R{\o}dsrud}},\ }\bibfield  {title} {\bibinfo {title} {Surface tension by
  pendant drop},\ }\href {https://doi.org/10.1016/0021-9797(91)90296-K}
  {\bibfield  {journal} {\bibinfo  {journal} {J.~Colloid~Interface~Sci}\
  }\textbf {\bibinfo {volume} {140}},\ \bibinfo {pages} {1} (\bibinfo {year}
  {1991})}\BibitemShut {NoStop}%
\bibitem [{\citenamefont {Eggers}\ and\ \citenamefont
  {Villermaux}(2008)}]{Eggers2008}%
  \BibitemOpen
  \bibfield  {author} {\bibinfo {author} {\bibfnamefont {J.}~\bibnamefont
  {Eggers}}\ and\ \bibinfo {author} {\bibfnamefont {E.}~\bibnamefont
  {Villermaux}},\ }\bibfield  {title} {\bibinfo {title} {Physics of liquid
  jets},\ }\href {https://doi.org/10.1088/0034-4885/71/3/036601} {\bibfield
  {journal} {\bibinfo  {journal} {Rep. Prog. Phys.}\ }\textbf {\bibinfo
  {volume} {71}},\ \bibinfo {pages} {036601} (\bibinfo {year}
  {2008})}\BibitemShut {NoStop}%
\bibitem [{\citenamefont {Day}\ \emph {et~al.}(1998)\citenamefont {Day},
  \citenamefont {Hinch},\ and\ \citenamefont {Lister}}]{Day1998}%
  \BibitemOpen
  \bibfield  {author} {\bibinfo {author} {\bibfnamefont {R.~F.}\ \bibnamefont
  {Day}}, \bibinfo {author} {\bibfnamefont {E.~J.}\ \bibnamefont {Hinch}},\
  and\ \bibinfo {author} {\bibfnamefont {J.~R.}\ \bibnamefont {Lister}},\
  }\bibfield  {title} {\bibinfo {title} {Self-similar capillary pinchoff of an
  inviscid fluid},\ }\href {https://doi.org/10.1103/PhysRevLett.80.704}
  {\bibfield  {journal} {\bibinfo  {journal} {Phys. Rev. Lett.}\ }\textbf
  {\bibinfo {volume} {80}},\ \bibinfo {pages} {704} (\bibinfo {year}
  {1998})}\BibitemShut {NoStop}%
\bibitem [{\citenamefont {Erneux}\ and\ \citenamefont
  {Davis}(1993)}]{Erneux1993}%
  \BibitemOpen
  \bibfield  {author} {\bibinfo {author} {\bibfnamefont {T.}~\bibnamefont
  {Erneux}}\ and\ \bibinfo {author} {\bibfnamefont {S.~H.}\ \bibnamefont
  {Davis}},\ }\bibfield  {title} {\bibinfo {title} {Nonlinear rupture of free
  films},\ }\href {https://doi.org/10.1063/1.858597} {\bibfield  {journal}
  {\bibinfo  {journal} {Phys. Fluids A}\ }\textbf {\bibinfo {volume} {5}},\
  \bibinfo {pages} {1117} (\bibinfo {year} {1993})}\BibitemShut {NoStop}%
\bibitem [{\citenamefont {Scheid}\ \emph {et~al.}(2012)\citenamefont {Scheid},
  \citenamefont {van Nierop},\ and\ \citenamefont {Stone}}]{Scheid2012}%
  \BibitemOpen
  \bibfield  {author} {\bibinfo {author} {\bibfnamefont {B.}~\bibnamefont
  {Scheid}}, \bibinfo {author} {\bibfnamefont {E.~A.}\ \bibnamefont {van
  Nierop}},\ and\ \bibinfo {author} {\bibfnamefont {H.~A.}\ \bibnamefont
  {Stone}},\ }\bibfield  {title} {\bibinfo {title} {Thermocapillary-assisted
  pulling of contact-free liquid films},\ }\href
  {https://doi.org/10.1063/1.3692097} {\bibfield  {journal} {\bibinfo
  {journal} {Phys. Fluids}\ }\textbf {\bibinfo {volume} {24}},\ \bibinfo
  {pages} {032107} (\bibinfo {year} {2012})}\BibitemShut {NoStop}%
\bibitem [{\citenamefont {Eggers}\ and\ \citenamefont
  {Fontelos}(2015)}]{Eggers2015}%
  \BibitemOpen
  \bibfield  {author} {\bibinfo {author} {\bibfnamefont {J.}~\bibnamefont
  {Eggers}}\ and\ \bibinfo {author} {\bibfnamefont {M.~A.}\ \bibnamefont
  {Fontelos}},\ }\href {https://doi.org/10.1017/CBO9781316161692} {\emph
  {\bibinfo {title} {Singularities: Formation, Structure, and Propagation}}}\
  (\bibinfo  {publisher} {Cambridge University Press},\ \bibinfo {year}
  {2015})\BibitemShut {NoStop}%
\bibitem [{\citenamefont {Ting}\ and\ \citenamefont {Keller}(1990)}]{Ting1990}%
  \BibitemOpen
  \bibfield  {author} {\bibinfo {author} {\bibfnamefont {L.}~\bibnamefont
  {Ting}}\ and\ \bibinfo {author} {\bibfnamefont {J.~B.}\ \bibnamefont
  {Keller}},\ }\bibfield  {title} {\bibinfo {title} {Slender jets and thin
  sheets with surface tension},\ }\href {https://doi.org/10.1137/0150090}
  {\bibfield  {journal} {\bibinfo  {journal} {SIAM J. Appl. Math.}\ }\textbf
  {\bibinfo {volume} {50}},\ \bibinfo {pages} {1533} (\bibinfo {year}
  {1990})}\BibitemShut {NoStop}%
\bibitem [{\citenamefont {Billingham}\ and\ \citenamefont
  {King}(2005)}]{Billingham2005}%
  \BibitemOpen
  \bibfield  {author} {\bibinfo {author} {\bibfnamefont {J.}~\bibnamefont
  {Billingham}}\ and\ \bibinfo {author} {\bibfnamefont {A.~C.}\ \bibnamefont
  {King}},\ }\bibfield  {title} {\bibinfo {title} {Surface-tension-driven flow
  outside a slender wedge with an application to the inviscid coalescence of
  drops},\ }\href {https://doi.org/10.1017/S0022112005004349} {\bibfield
  {journal} {\bibinfo  {journal} {J. Fluid Mech.}\ }\textbf {\bibinfo {volume}
  {533}},\ \bibinfo {pages} {193} (\bibinfo {year} {2005})}\BibitemShut
  {NoStop}%
\bibitem [{\citenamefont {Eggers}(1993)}]{Eggers1993}%
  \BibitemOpen
  \bibfield  {author} {\bibinfo {author} {\bibfnamefont {J.}~\bibnamefont
  {Eggers}},\ }\bibfield  {title} {\bibinfo {title} {Universal pinching of 3d
  axisymmetric free-surface flow},\ }\href
  {https://doi.org/10.1103/PhysRevLett.71.3458} {\bibfield  {journal} {\bibinfo
   {journal} {Phys. Rev. Lett.}\ }\textbf {\bibinfo {volume} {71}},\ \bibinfo
  {pages} {3458} (\bibinfo {year} {1993})}\BibitemShut {NoStop}%
\bibitem [{\citenamefont {Laan}\ \emph {et~al.}(2014)\citenamefont {Laan},
  \citenamefont {de~Bruin}, \citenamefont {Bartolo}, \citenamefont
  {Josserand},\ and\ \citenamefont {Bonn}}]{Laan2014}%
  \BibitemOpen
  \bibfield  {author} {\bibinfo {author} {\bibfnamefont {N.}~\bibnamefont
  {Laan}}, \bibinfo {author} {\bibfnamefont {K.~G.}\ \bibnamefont {de~Bruin}},
  \bibinfo {author} {\bibfnamefont {D.}~\bibnamefont {Bartolo}}, \bibinfo
  {author} {\bibfnamefont {C.}~\bibnamefont {Josserand}},\ and\ \bibinfo
  {author} {\bibfnamefont {D.}~\bibnamefont {Bonn}},\ }\bibfield  {title}
  {\bibinfo {title} {Maximum diameter of impacting liquid droplets},\ }\href
  {https://doi.org/10.1103/PhysRevApplied.2.044018} {\bibfield  {journal}
  {\bibinfo  {journal} {Phys. Rev. Appl.}\ }\textbf {\bibinfo {volume} {2}},\
  \bibinfo {pages} {044018} (\bibinfo {year} {2014})}\BibitemShut {NoStop}%
\bibitem [{\citenamefont {Paulsen}\ \emph {et~al.}(2014)\citenamefont
  {Paulsen}, \citenamefont {Carmigniani}, \citenamefont {Kannan}, \citenamefont
  {Burton},\ and\ \citenamefont {Nagel}}]{Paulsen2014}%
  \BibitemOpen
  \bibfield  {author} {\bibinfo {author} {\bibfnamefont {J.~D.}\ \bibnamefont
  {Paulsen}}, \bibinfo {author} {\bibfnamefont {R.}~\bibnamefont
  {Carmigniani}}, \bibinfo {author} {\bibfnamefont {A.}~\bibnamefont {Kannan}},
  \bibinfo {author} {\bibfnamefont {J.~C.}\ \bibnamefont {Burton}},\ and\
  \bibinfo {author} {\bibfnamefont {S.~R.}\ \bibnamefont {Nagel}},\ }\bibfield
  {title} {\bibinfo {title} {Coalescence off bubbles and drops in an outer
  fluid},\ }\href {https://doi.org/10.1038/ncomms4182} {\bibfield  {journal}
  {\bibinfo  {journal} {Nat. Commun.}\ }\textbf {\bibinfo {volume} {5}},\
  \bibinfo {pages} {3181} (\bibinfo {year} {2014})}\BibitemShut {NoStop}%
\end{thebibliography}
\end{document}